\documentclass[]{ceurart}

\usepackage{booktabs}
\usepackage{graphicx}

\begin{document}

\copyrightyear{2026}
\copyrightclause{Copyright for this paper by its authors.
  Use permitted under Creative Commons License Attribution 4.0
  International (CC BY 4.0).}

\conference{PEAF 2026: Workshop on Pedagogical Evaluation of Automated Feedback, June 28, 2026, Seoul, South Korea}

\title{The Correct Answer Trap: Pedagogically-Grounded Detection and Feedback for Hidden Misconceptions}

\author[1]{Moiz Imran}[%
email=moiz.imran.22@ucl.ac.uk
]  

\author[1,2]{Sahan Bulathwela}[%
orcid=0000-0002-5878-2143,
email=m.bulathwela@ucl.ac.uk
 ]

\address[1]{Department of Computer Science, University College London, The United Kingdom}
\address[2]{Centre for Artificial Intelligence, University College London, The United Kingdom}

\begin{abstract}
Automated feedback systems that rely on answer correctness will reinforce, rather than address, misconceptions when students reach the correct answer through flawed reasoning. We investigate automatic detection of these hidden misconceptions using 20,964 real student responses from the Eedi mathematics platform. Fine-tuned classifiers detect only 57\% of these hidden misconceptions, and standard ML interventions do not improve on this. An open-weight reasoning model detects 84\%, but at realistic prevalence, false alarms outnumber genuine detections roughly 8 to 1. We present a graduated assessment rubric that separates answer correctness from method validity, and propose a detect-verify-escalate pipeline that routes uncertain cases to diagnostic follow-up questions rather than directly to teachers. Two deployment modes adapt the pipeline: a teacher dashboard where the system filters a review queue, and an autonomous tutor where flags trigger low-cost formative follow-up.
\end{abstract}

\begin{keywords}
 Misconception Detection \sep
 Automated Feedback \sep
 Formative Assessment \sep
 LLM \sep
 Pedagogical Evaluation
\end{keywords}

\maketitle

\section{Introduction}

Intelligent tutoring systems rely on automated assessment to decide what feedback a student should receive. For this feedback to be useful, the system requires evidence about the student's reasoning, not merely the final answer. Short written explanations can fill this gap, provided models can interpret them reliably.

Consider a student solving $(-8) - (-5)$ who writes: ``I did 8 minus 5, which is 3, and since there are negatives the answer is $-3$.'' The answer is correct, but the reasoning applies a flawed rule: subtract the smaller absolute value from the larger and assign a negative sign. On $(-5) - (-8)$, the same procedure gives $8 - 5 = 3$, then $-3$, but the correct answer is $+3$. If the system checks only the answer, it confirms a misconception it should be correcting.

We term this the \emph{correct answer trap} (CAT): the under-detection of misconceptions when flawed reasoning produces a correct answer \cite{imran2026catching}. For automated feedback systems, these cases are the hardest setting because the standard signal of a mistake (a wrong answer) is absent, and detection depends entirely on interpreting the student's explanation.

Building on our characterisation of this problem \cite{imran2026catching}, we show why models fail at this task (fine-tuned classifiers default to answer correctness as a shortcut, and specific questions create paths where flawed reasoning produces correct outputs). We then present a graduated assessment formulation grounded in the mark/method distinction from educational assessment, and a detect-verify-escalate architecture that works around these failure modes to connect detection to pedagogically differentiated feedback.

\section{Background}

Where recent work in AI tutoring evaluates the reception and outcomes of generated feedback \cite{daheim2024stepwise}, we address the prior question: how should a system decide what feedback to give when correctness, reasoning quality, and diagnostic certainty diverge? Black and Wiliam's formative assessment framework \cite{black1998assessment} identifies three steps: elicit evidence of learning, interpret that evidence, and respond appropriately. The correct answer trap is a failure at the interpretation step: the evidence is misinterpreted as understanding, so the response (confirmation) is pedagogically harmful.

Barton \cite{barton2018diagnosticqs} identifies a principle of diagnostic question design: ``it should not be possible to answer the question correctly whilst still holding a key misconception.'' When this principle is violated, students applying flawed procedures can coincidentally produce correct answers, a pattern Kapur \cite{kapur2016examining} terms ``unproductive success'' and Brown and Burton \cite{brownburton1978} formalised as procedural bugs that produce correct outputs for specific numerical inputs. The two questions that account for 71\% of hidden misconceptions in our dataset both violate this principle: their structure allows flawed procedures to produce correct outputs. The largest benchmark for AI misconception detection, the Eedi Kaggle competition \cite{kaggle_eedi2024}, tasks models with predicting which misconception explains a \emph{wrong} answer, excluding correct-answer misconceptions by construction. We address the distinct \emph{detection} problem: determining whether a misconception exists at all, even when the answer is correct.

\section{Method}

\subsection{Data and Task}

We use the Eedi dataset: 20,964 real student responses to 15 diagnostic mathematics questions whose incorrect options are each mapped to a specific misconception. Each response consists of the question, student's MCQ answer, and a short written explanation (average 16 words). Responses are labelled as True Correct (TC, 60\%), False Misconception (FM, 38\%), or True Misconception (TM, 1.6\%). TM cases are the correct answer trap: correct answer, flawed reasoning. We evaluate how language models can detect TMs on a held-out test set of 3,702 responses containing 61 TM cases.

\subsection{Graduated Assessment Design}

Standard mathematical assessment criteria distinguish between the \emph{answer mark} (correct output) and the \emph{method mark} (valid reasoning process). A student can earn the answer mark without the method mark. Wrong answers are identified by the answer key alone. For correct-answer cases, the method can be clearly valid, clearly flawed, or indeterminate because the explanation is too brief to judge. We use this distinction to motivate a graduated rubric with four response categories:

\begin{enumerate}
  \item \textbf{Clear reasoning.} Correct answer, valid method.
  \item \textbf{Needs clarification.} Correct answer, but explanation too brief to judge.
  \item \textbf{Misconception detected.} Correct answer, flawed reasoning visible.
  \item \textbf{Wrong answer.} Incorrect option selected.
\end{enumerate}

A binary ``misconception or not'' formulation collapses Options 2 and 3, either falsely accusing students whose explanations are merely brief, or losing the routing signal that distinguishes unclear from flawed reasoning.

The model receives the question, student answer, explanation, and a mark scheme describing the expected reasoning method. Testing three prompt variants on the same model and test set, we find performance robust to minor wording changes (1 percentage point (pp) balanced accuracy difference) but sensitive to holistic rubric redesign (27pp drop). This confirms that the rubric structure is part of the assessment construct \cite{messick1989validity}: what the task asks the model to distinguish matters more than how the prompt phrases it.

\subsection{Models}

We compare three model classes: (1) BERT-base fine-tuned on the training split with weighted cross-entropy \cite{devlin2019bert}, representing fine-tuned classifiers; (2) Gemini 3 Flash, a frontier-level proprietary model selected as the most cost-efficient reasoning-capable LLM at time of evaluation; and (3) Gemma 4 26B, an open-weight model whose active parameter count (3.8B) enables deployment on consumer hardware while achieving performance competitive with substantially larger models \cite{gemmateam2026gemma4}. All LLMs use the graduated assessment with the mark scheme prompt.

\section{Results}

Table~\ref{tab:results} reports detection performance across the three model classes. BERT detects only 57\% of correct-answer misconceptions despite near-perfect wrong-answer classification (FM recall 100\%). Several standard ML interventions (class-weighted sampling, data augmentation, focal loss, cross-encoder architecture, and NLI reformulation) all fall within the baseline confidence interval. Integrated gradients analysis confirms the mechanism: misclassified cases show answer-dominant attribution, while correctly classified cases show explanation-dominant attribution \cite{geirhos2020shortcut}.

\begin{table}[t]
\centering
\small
\caption{Graduated assessment results on the test set ($n = 3{,}702$). TM Det.\ = misconception detected; TM Flag.\ = needs clarification; TC FP = true correct false positive rate (correct students wrongly flagged). Wilson 95\% CIs on TM detection.}
\label{tab:results}
\begin{tabular}{@{}l cc cc | c@{}}
\toprule
 & \multicolumn{2}{c}{TM ($n=61$)} & \multicolumn{2}{c}{TC ($n=2{,}221$)} & \\
\cmidrule(lr){2-3} \cmidrule(lr){4-5}
Model & Det.\ [\small{95\% CI}] & Flag. & Correct & FP & FM Recall \\
\midrule
BERT baseline & 57.4 [44.5, 69.4] & --- & 99.3 & 0.7 & 100 \\
Gemini 3 Flash & 70.5 [57.7, 80.9] & 18.0 & 74.1 & 8.9 & 83.0 \\
Gemma 4 26B & 83.6 [72.1, 91.4] & 11.5 & 62.4 & 18.0 & 87.8 \\
\bottomrule
\end{tabular}
\end{table}

Reasoning-capable models clear this threshold: Gemma 4 detects 84\% and flags a further 12\% for clarification. However, at 1.6\% prevalence, even 84\% detection with 18\% false positive rate yields only 10.9\% positive predictive value (roughly 8 false alarms per genuine detection). Standalone binary classification is not viable at any foreseeable accuracy; a teacher in the loop is a necessity.

An ablation removing the mark scheme from Gemma 4's prompt leaves TM detection unchanged but raises the TC false positive rate from 18.0\% to 25.4\% ($p < 0.001$). The mark scheme does not help the model \emph{detect} misconceptions; it prevents unnecessary false positives against students with correct reasoning.

\section{The Feedback Pipeline}

\begin{figure}
\centering
\includegraphics[width=0.55\linewidth]{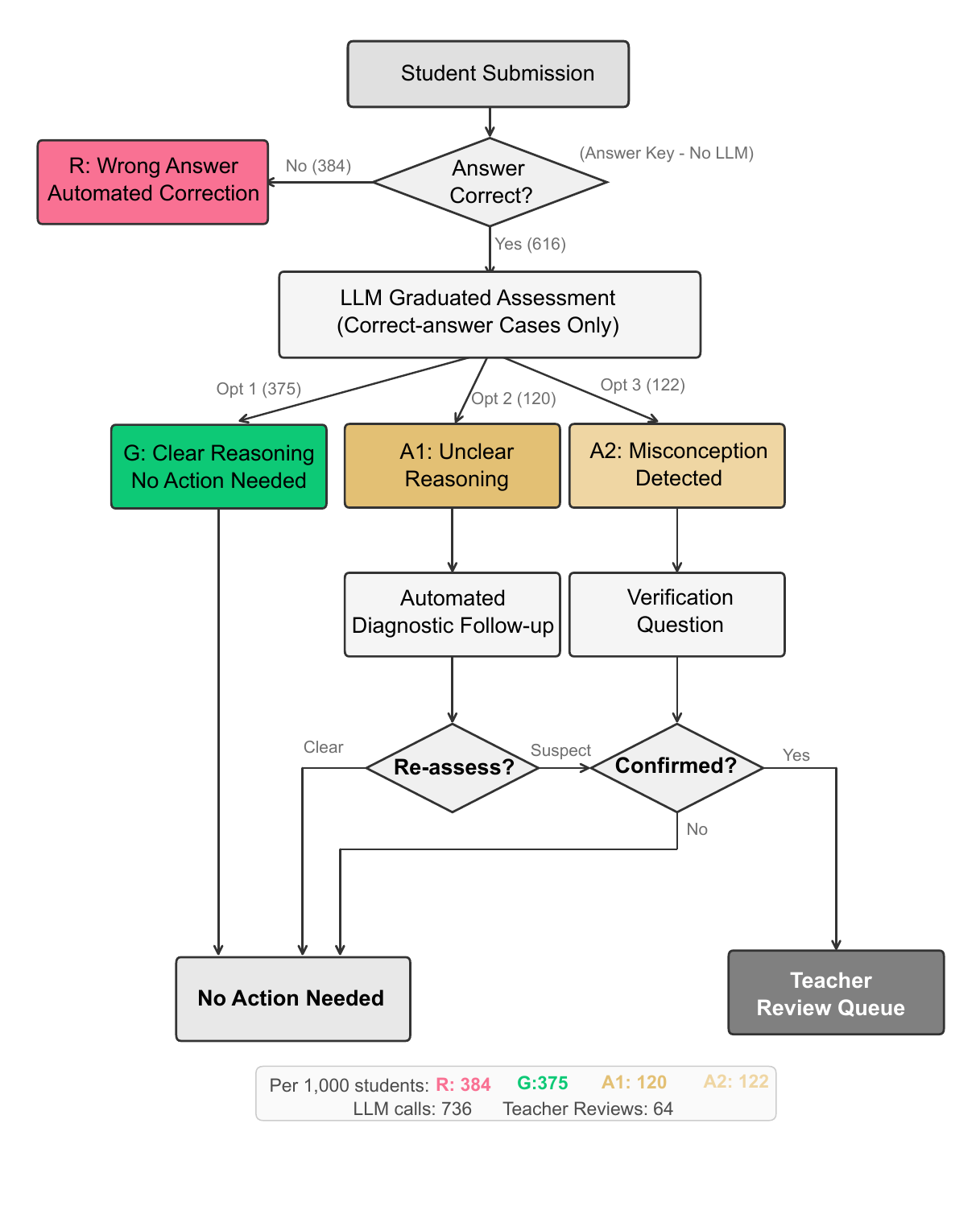}
\caption{The detect-verify-escalate pipeline. Wrong answers (Red) are filtered by the answer key. Correct-answer cases are assessed using the graduated rubric, producing three routes: Clear Correct Reasoning (Green), Unclear Reasoning (Amber 1) and Misconception Detected (Amber 2). Approximate volumes per 1,000 submissions are shown.}
\label{fig:architecture}
\end{figure}

These findings jointly constrain what a viable feedback architecture needs: reasoning capability (fine-tuned models cannot assess method validity), graduated output (binary classification loses the routing signal), and intermediate verification (at 1.6\% prevalence, precision is too low for direct teacher escalation). Figure~\ref{fig:architecture} shows the resulting detect-verify-escalate pipeline with four routes, each mapping to a pedagogically distinct response \cite{black1998assessment}: Red (wrong answer, handled by answer-key lookup), Green (clear reasoning, confirmation), Amber-1 (needs clarification, follow-up question), and Amber-2 (misconception detected, verification then teacher escalation).

The Amber-1/Amber-2 distinction is pedagogically meaningful. Of the TM cases Gemma 4 routed to Option 2, the majority carry the ``Irrelevant'' annotation code: human annotators also could not identify a specific misconception. For these cases, the same model generates a targeted diagnostic follow-up question using either a \emph{variant-problem strategy} \cite{barton2018diagnosticqs} (presenting a near-transfer problem where the suspected misconception would produce the wrong answer) or a \emph{probe-the-gap strategy} (asking about the specific step the student omitted).

We propose two deployment configurations for future implementation. In a \textbf{teacher dashboard} mode, the system would filter and present a review queue, so a conservative model suits this context. In an \textbf{autonomous tutor} mode, each flag would trigger a follow-up question. Since asking a student to explain their reasoning is good formative practice regardless of whether a misconception is present \cite{black1998assessment}, false positives are pedagogically benign in this mode; false negatives (student leaves with misconception unchallenged) are the expensive error.

\section{Discussion}

Our results show that fine-tuned classifiers miss over 40\% of correct-answer misconceptions, and that reasoning-capable models reduce this gap but introduce false positive rates that make standalone classification impractical at classroom prevalence. These findings have implications for how automated feedback systems are evaluated. If the diagnostic routing is wrong, the pedagogical quality of any subsequent feedback is irrelevant. Our work shows that correct-answer cases, where the standard error signal is absent, require dedicated evaluation of the detection step itself.

The graduated rubric addresses a related problem. Short student explanations (average 16 words) often contain genuinely ambiguous evidence. Rather than forcing a binary misconception decision on insufficient information, the rubric allows the system to express uncertainty and gather further evidence through targeted follow-up. This treats diagnostic ambiguity as a property of the evidence rather than a failure of the model. In the context of lifelong learning \cite{bulathwela2025truereason} and Knowledge Tracing-based ITS systems \cite{norris2025ntkt}, the proposed misconception detection system can i) allow better recovery of the student state and ii) provide precise feedback, driven by the misconceptions. 

\subsection{Limitations and Future Work}

All findings come from 15 mathematics questions on one platform, with TM concentrated in 2 items. The diagnostic follow-up proof of concept generates follow-up questions for 7 TM instances from the test set; no student responses to these questions have been collected. A classroom trial with real students answering the generated follow-up questions would validate whether the pipeline reduces teacher workload while catching misconceptions that generous marking lets through.

\begin{acknowledgments}
This work is co-funded by the European Commission's projects ``Teacher-AI Complementarity (TaiCo)'' (Project ID: 101177268), ``Humane AI'' (Grant No.\ 820437) and ``X5GON'' (Grant No.\ 761758) and the UCL Computer Science Strategic Research Fund.
\end{acknowledgments}

\section*{Declaration on Generative AI}
During the preparation of this work, the authors used Claude (Anthropic) for drafting content, grammar checking, and spelling checking. After using these tools, the authors reviewed and edited the content as needed and take full responsibility for the publication's content.

\bibliography{references}

\end{document}